# Skirting Terahertz Waves in a Photo-excited Nanoslit Structure

*Mostafa Shalaby[1]\*, Justyna Fabiańska[2], Marco Peccianti[1,3], Yavuz Ozturk[1], Francois Vidal[1], Hans Sigg[4], Roberto Morandotti[1], and Thomas Feurer[2]*

[1]INRS-EMT, Varennes, Quebec, J3X 1S2, Canada

[2]Institute of Applied Physics, University of Bern, Sidlerstrasse 5, CH-3012 Bern, Switzerland

[3]Also at: Dept. of Physics and Astronomy, University of Sussex, Falmer, Brighton BN1 9QH, United Kingdom

[4]LMN, Paul Scherrer Institut, Villigen 5232, Switzerland

Terahertz fields can be dramatically enhanced as they propagate through nanometer-sized slits. The enhancement is mediated by significant accumulation of the induced surface charges on the surrounding metal. This enhancement is shown here to be dynamically modulated and the nanoslits are gradually shunted using a copropagating optical beam. The terahertz fields are found to skirt the nanoscale photo-excited region underneath the slits, scattering to the far field and rigorously mapping the near field.

The observation of the extraordinary optical transmission (EOT)[1] phenomenon sparked the research of light control in sub-wavelength structures[2-15]. The increasing fundamental understanding of the underlying mechanisms keeps adding new applications in the linear (e.g. molecular spectroscopy and sensing[16-20]) and nonlinear regimes. Nonlinear applications are based on the local field enhancement (FE) associated with the EOT. The two phenomena are usually interconnected. On one hand, an electromagnetic wave incident on a resonant sub-wavelength structure can efficiently couple to the surface plasmon polaritons (SPP) leading to EOT[21-22], which results in high FE in the apertures[23-28]. On the other hand, a single infinitely long narrow slit was shown to lead to a high concentration of field on the sides of the slit, which results in a broadband EOT[29-30].

A slit etched in a thin metal film can be depicted as a nanocapacitor charged by the light-induced charges on the surrounding metal surface[29]. The charge concentration increases as the gap width (a) decreases and the wavelength ($\lambda$) increases, leading to extreme FE on the nanoscale[29-31]. This gives nanoslits great potential for the (long) terahertz wavelengths (e.g. reports on FE = 720 at 0.1 THz in 70 nm-wide single slits[29] and FE = 760 at 0.2 THz in 40 nm-wide slits arrays[30]).

Advances in terahertz technology have recently boosted critical applications in the linear regime, such as spectroscopy of chemicals and explosives[32-33], remote sensing[34], communications[35], and imaging[36-39]. On the nonlinear side, although there have been a few reports[40-43], the technology of terahertz sources[44-46] and systems[47-48] still lags behind the demand of many awaiting experiments[49-51].

Field enhancement is a linear phenomenon that depends on the structure properties and dielectric environment at the operating frequency. Dynamic control over the FE is highly desired in many nonlinear applications. A possible approach is to locally change the dielectric environment. As semiconductors are typical substrate materials, optical excitation could in principle significantly change the substrate conductivity and dielectric environment, and thus modulate the FE and the transmission (T). Here we exploit this property to control the local FE and propagation of the terahertz (1 THz corresponds to 300 μm) waves using optical fields (800 nm, i.e. more than 2 orders of magnitude shorter in wavelength) of a 40 nm-wide slit deposited on a silicon substrate. The whole process (optical excitation/terahertz enhancement modulation) effectively occurs on the nanoscale around the slit. In particular, this work demonstrates that the photo-induced carriers in the underlying substrate shunt the slit (hence modulating its field enhancement and consequently its transmission) as it creates an Ohmic contact between the two metal sidewalls. The simple THz absorption in the substrate plays a much less significant role. In this process, there is an important issue related to the practical estimation of the enhanced field. As it is not yet possible to directly access the local field inside the slit, the correlation between the measured far field, the near field and FE modulation will be hereby discussed.

In a recent paper, we have demonstrated how the normally very low THz transmission from a nanoslit can be significantly enhanced using a linear array of nanoslits. In light of the limited sensitivity of a standard time domain spectroscopy, this design is particularly suited for investigating the modulation of both T and FE[30]. This structure consists of an array of 40 nm-wide 100 μm-spaced slits, etched in a thin (60 nm-thick) gold film. The slits are 2 mm-long and we have a total of 20 slits on the sample. As a substrate, we used a high resistivity 500 μm-thick silicon (Fig. 1). We performed the experimental measurements using an optical pump/terahertz probe configuration. The laser pulses (energy = 2 mJ, duration = 130 fs, repetition rate = 1 kHz, center wavelength = 800 nm) were split between the optical pump and the terahertz probe branches. The latter is then split between the generation (using optical rectification) and detection (using electro-optical sampling) in two different ZnTe crystals. The pump/probe delay was 10 ps, and thus much longer than the pump pulse duration and much shorter than the carrier lifetime in all the measurements. The pump/probe configuration at the sample place is shown in Fig. 1 where the sample is placed orthogonally to the terahertz propagation axis. The optical pump impinges on the sample collimated at an incidence angle of 30°.

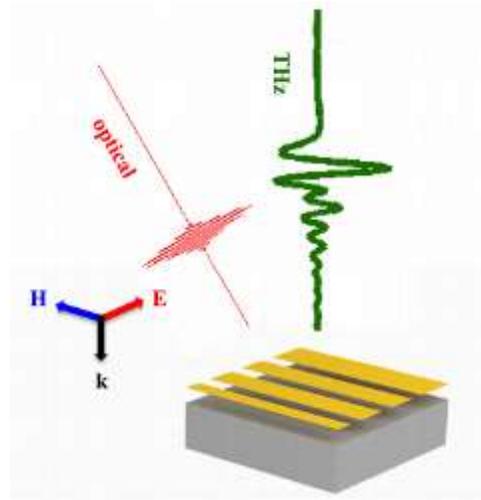

**Figure 1.** Schematic diagram of the pump-probe configuration and the nanoslits orientation with respect to the polarization of the fields ($E = E_x$, $H = H_y$, $k = k_z$). The incidence angle of the optical pump is 30°. The geometrical parameters of the nanoslit array are: slit width 40 nm, periodicity 100 μm, gold film thickness 60 nm, and slit length 2 mm.

Narrow (sub-wavelength) slits show a strongly polarization-dependent response. Their FE and EOT properties are exploited when the electric field of the light is polarized along the slit width (that is, $E = E_x$). This condition was ensured for both the optical and terahertz pulses in all the measurements presented here. Our study is concerned with the interaction between the slit structure and the two beams, the optical (O) and the terahertz (THz) with the respective wavelengths $\lambda_o$ and $\lambda_{THz}$. In principle, a $\ll \lambda_o$ & $\lambda_{THz}$, so interesting transmission and field-enhancement properties are expected in both regimes. However, the general physical settings at the two wavelengths are quite different. The two main mechanisms of interest here are the field concentration around metal edges and the surface plasmons. First, an electric field impinging on a conducting surface induces surface currents, which tend to accumulate charges around the edges (the gap). This effect scales as $\sqrt{\lambda}$ and is thus much more pronounced in the terahertz case[29]. Second, surface plasmons -which are the surface coupled modes at the metal's surface- are supported (that is, localized at the metal surface) at the optical frequencies (but with the resonance outside the bandwidth of our optical pulse) and work to enhance the optical waves close to the metal surface. Conversely, those modes at the terahertz frequencies are delocalized and radiate away. This is due to the large real conductivity at terahertz frequencies, which does not allow for significant terahertz penetration inside the metal[52-53].

Our discussion here is focused on the terahertz electric field and optical power. In the present configuration, the terahertz magnetic field is parallel to the slit direction and is thus barely enhanced. It just spreads out in an almost homogenous fashion. Regarding the optical excitation, the role of the pump is to change the dielectric properties of the substrate upon the absorption of optical photons. This change modulates the propagation of a delayed terahertz pulse through the structure. It is then important to assess the distribution of the power flow (**P**) underneath the slit plane. As the slit length is 2 mm $\gg \lambda_{\text{THz}}$, we assume no dynamics along the (y-) slit axis and so we restrict our plots to the x-z plane. Figure 2a shows a 2D map of |**P**| under the slits plane. For a given input power $\mathbf{P_0}$, after accounting for the oblique incidence, **P** in the substrate beneath the slit is comparable to $\mathbf{P_0}$, *i.e.* there is no significant optical power enhancement (away from the slit edges). Rather, the structured film just screens the input waves. Inside the substrate, the optical field is absorbed with an absorption length on the order of 16 $\mu$m since the photon energy (1.55 eV) is larger than the band gap (1.11 eV).

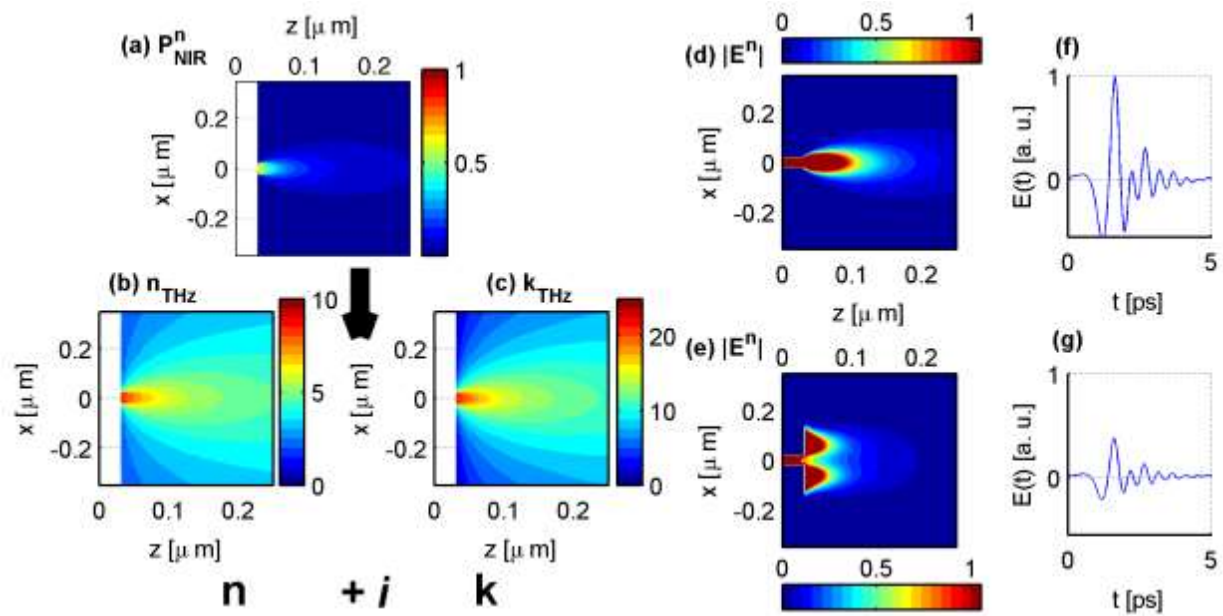

**Figure 2.** (a) The optical power distribution under the slit (at a fluence of 900 □J/cm$^2$). Both the optical pump and the THz probe are incident from the left. This excitation corresponds to a modified terahertz dielectric response of the substrate: (b) real and (c) imaginary parts of the modified refractive index at 0.93 THz. (d) The x-component of the terahertz electric field distribution around the slit under no optical excitation with the corresponding far field time domain spectroscopy measurement shown in (f). (e) The effect of optical excitation on the near field distribution of a 10 ps-delayed terahertz pulse. (g) Measurement of the corresponding terahertz pulse in the far field, showing attenuation compared to that in (f). (a), (d),(e) are normalized quantities. (b),(c),(d),(e) are simulated at 0.93 THz.

Absorption of optical photons in the semiconductor substrate leads to the creation of free carriers $N(\text{x}, \text{z})$, which translates into the plasma frequency $\omega_{\text{p}}(\text{x}, \text{z}) = N(\text{x}, \text{z})e^2/\varepsilon_0 m^*$ where $e, \varepsilon_0$, and $m^*$ are the electron charge, free space permittivity, and electron effective mass, respectively. The increasing plasma frequency leads to a change in the semiconductor's dielectric constant $\Delta\varepsilon = \varepsilon - \varepsilon_\infty = \omega_{\text{p}}^2/(\omega^2 - i\gamma\omega)$ with $\varepsilon_\infty$ and $\gamma$ being the dielectric constant of unexcited silicon and the collision frequency, respectively. In order to numerically reproduce our experiment, we used the following recipe: i) calculate a 2D (x-z) map of the optical power distribution (at a given fluence) under the slit (Fig. 2a), ii) convert this power map into a complex $\varepsilon$ for the terahertz frequency or, as shown here, for the refractive index n $+ i$ k map and use it to update the substrate's dielectric function in the respective region (Fig. 2b,c), and iii) simulate the terahertz propagation through the modified dielectric environment (Fig. 2d,e) and compare the simulated far fields with our time domain spectroscopy measurements (Fig. 2f,g).

Details on the numerical calculations are given in the supporting materials. The simulations in Fig. 2 are shown at a frequency of 0.93 THz and an optical fluence of 900 μJ/cm². As a consequence of the optical pumping, the THz refractive index evolves from its equilibrium real value (3.418) to a complex quantity in the region beneath the gap. In a typical far field measurement, this translates into a reduced transmission as can be seen by comparing Fig. 2g with Fig. 2f.

"What is the underlying mechanism behind the far field attenuation?" is a fundamental question to answer. In principle, there are two attenuation scenarios that can take place here. In the first, the losses (absorption and reflection) in the photo-excited region account for the reduced transmission. This implies that i) the optical pump will not change the terahertz propagation direction after the slits plane and ii) the enhancement inside the slit will not change (as it takes place before the terahertz gets absorbed and/or reflected). The second attenuation scenario explains the reduced transmission in terms of a change in the slit enhancement mechanism and even shunting the slit gap. In Ref. 29, Seo et al. describe a nanoslits narrower than skin depth as a charged capacitor with a strong force across the gap of its two oppositely-charged sides. As the number of free carriers in the photo-excited substrate increases, this capacitor-like effect fades away and the terahertz-induced charges no longer flow towards the gap and accumulate there. This means - in comparison with the first scenario - that the terahertz enhancement decreases and the terahertz waves simply skirt the excited region. Figure 2e demonstrates how the terahertz goes around the photo-excited regions and scatters towards the far field. This confirms that the second mechanism is more relevant to our experiment. In order to further illustrate the mechanism, we show in Fig. 3 the change in terahertz field distribution as the photo-excited region builds up. Considering the electric field $E_x$, as the optical fluence increases, the front of the emerging terahertz waves starts to split at the photo-excited region. This is mediated by a high concentration of emerging terahertz field in small "pockets" under the corners of the slits. Those regions, surrounded by the film plane and curved side of the excited region, represent the only places the terahertz can "sink" into from the nanogap. As they are tapered and small, the electric field $E$ goes up significantly in comparison with the unexcited state $E^0$. We now consider the field distribution in the substrate along the slit central line. Right after the slit output plane, the terahertz field drops significantly and even vanishes completely at high fluences. In the direction of propagation, simulations show that this region extends up to a few hundred nm, which is much less than the skin depth of 800 nm in silicon (~16 $\mu$m). This is due to the fast decay of the optical concentration beyond the slit plane (Fig. 2a). It also suggests that using other substrates with a much narrower skin depth will barely affect that extension. Rather, it will affect the photo-induced conductivity and the dielectric properties near the slit (that is the complex refractive index scale in Fig. 2b,c. The relative transmission ($t = E/E^0$) after this region is slightly different from that inside the slit. This confirms the earlier prediction that the terahertz waves simply go around the photo-excited region rather than be absorbed into it.

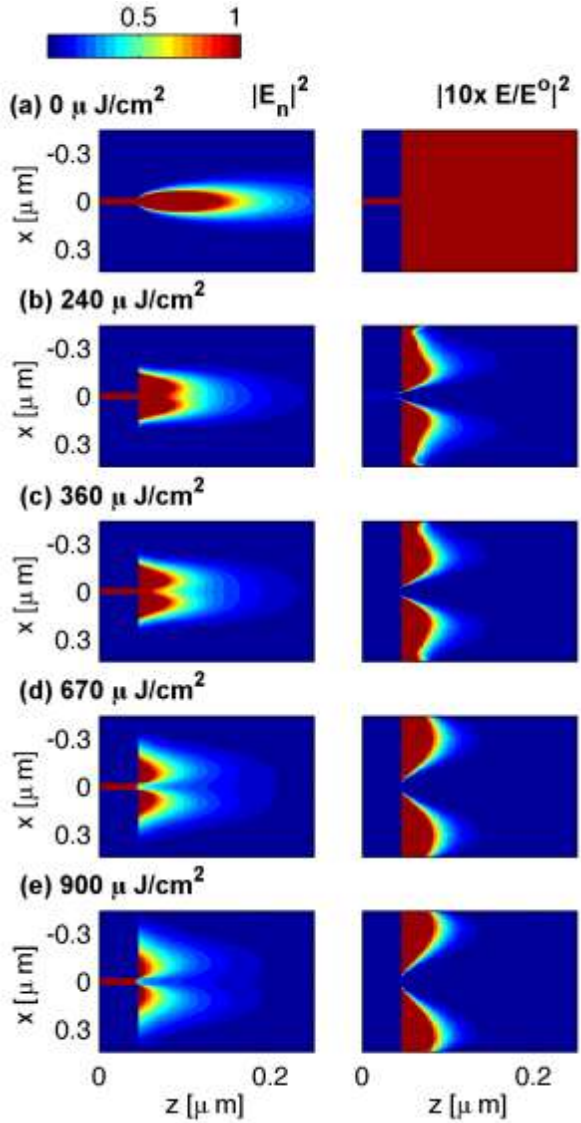

**Figure 3.** Absolute square of the terahertz electric field and of the relative terahertz field, shown for different levels of fluence. These simulations were performed at 0.93 THz. $E_n$ refers to the electric field normalized by the field in the center of the slit. $E$ and $E^0$ refer to the electric field in the presence and absence of the optical excitation respectively.

We conclude our work by discussing the effect of the optical pump on the FE and the far field-near field relations. As highlighted above, optical pumping of the nanoslit structure leads to a reduction of the transmitted terahertz field. This is depicted in the far field measurement (Fig. 4a). In Fig. 4b, we show the relative transmission in the 0.7-2 THz range. It is slightly frequency dependent and a minimum of 30% is obtained at 1 THz. This corresponds to a maximum modulation $\left(\frac{|E^0|^2 - |E|^2}{|E^0|^2}\right)$ of 91% under a fluence of 900 μJ/cm². The numerical calculations of the transmitted field at some frequency points are compared with the experimental measurements (Fig. 4b), showing good agreement. The recovery of the terahertz field, shown in Fig. 3, suggests that the unexcited far field/near field ratio holds even under optical excitation (note that without optical excitation, in this kind of structure, this ratio is given by the geometrical filling factor of the structure $\beta = 100 \, \mu m/40 \, nm = 2500$ and the enhancement factor $FE = t \times \beta$). To verify that, in Fig. 4c we show the variation in FE (at 1.25 THz, referenced to the unpumped sample) with optical fluence, obtained by way of two techniques. In the first, the simulated field is directly probed in the center of the slit. In the second, we scale the far field with $\beta$. The agreement between them confirms our prediction that the terahertz absorption in the photo-excited region is minimal and the dominant mechanism for reduced transmission is indeed enhancement modulation. As the enhancement is still linearly related to transmission, the maximum modulation in enhancement obtained here is 91% as well. We would like to highlight that, to the best of our knowledge, there is no experimental technique that is able to probe the relevant THz polarization inside such a narrow slit without disturbing the near field.

In conclusion, we have shown that the extreme field enhancement in nanoslits on a semiconductor substrate can be dynamically tuned using optical excitation. This leads to the creation of a photo-induced free carrier region where the terahertz field penetration is strongly reduced. The terahertz waves are, however, found to *skirt* this excited region and scatter to the far field, preserving the far field/near field relation. We believe that this work will help understand the behavior of these nanostructures in both optical and terahertz regimes, as well as in designing and interpreting some of the measurements in nonlinear experiments.

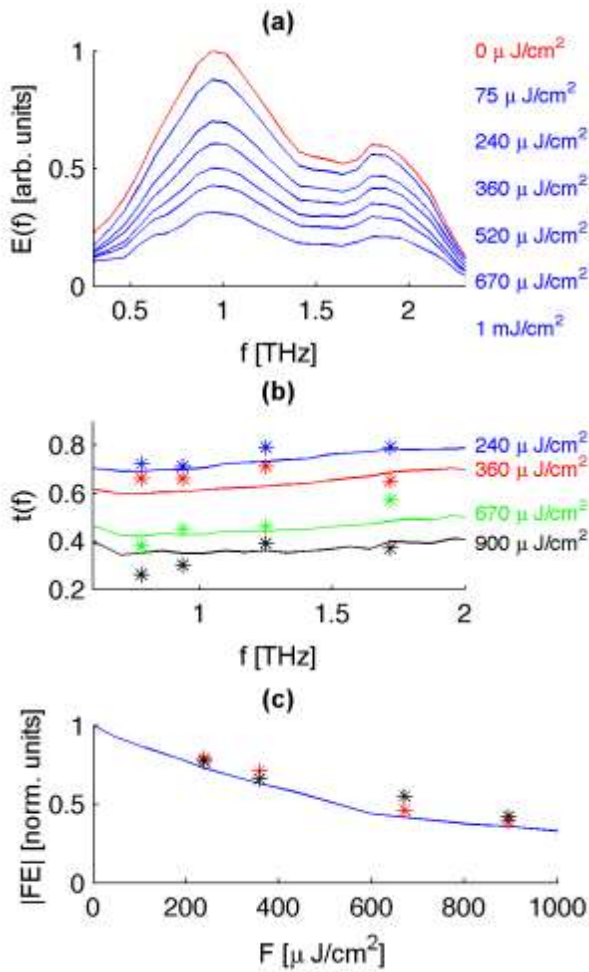

**Figure 4.** (a) The spectral amplitude of the transmitted terahertz far-fields at different fluence levels of the optical pump pulse. (b) The transmission spectra normalized to the unpumped case in the far field obtained from experiment (solid lines) and frequency domain simulations (stars). (c) Variation of the (normalized) terahertz field enhancement with the fluence probed at 1.25 THz. The solid curve shows the enhancement obtained by scaling the far field measurement with the structure filling factor. The stars denote the results obtained from frequency domain simulations by directly probing the near field (black) and scaling the far field by the filling factor (red).

**Corresponding Author**

*shalaby@emt.inrs.ca

ACKNOWLEDGMENT

We would like to gratefully thank L. J. Heyderman and A. Weber (ETH Zurich) for fabricating the structure. This work is supported by the Canadian FQRNT and NSERC, and the Swiss SNSF NCCR MUST funding agencies. M.S. wishes to acknowledge a FQRNT (MELS) scholarship.